\documentclass[aps,prb,showpacs,superscriptaddress,twocolumn]{revtex4}
\usepackage{amsmath}
\usepackage{amsfonts}
\usepackage{graphicx}
\usepackage{overpic}
\usepackage{subfigure}
\usepackage{natbib}
\usepackage{epstopdf}

\begin{document}

\title{Quantum walk on the line with quantum rings}
\author{Orsolya K\'{a}lm\'{a}n}
\affiliation{Department of Quantum Optics and Quantum Information, Research 
Institute for Solid
State Physics and Optics,Hungarian Academy of Sciences, 
Konkoly-Thege Mikl\'{o}s \'{u}t 29-33, H-1121 Budapest, Hungary}
\author{Tam\'{a}s Kiss}
\affiliation{Department of Quantum Optics and Quantum Information, Research 
Institute for Solid State Physics and Optics,Hungarian Academy of Sciences, 
Konkoly-Thege Mikl\'{o}s \'{u}t 29-33, H-1121 Budapest, Hungary}
\author{P\'{e}ter F\"{o}ldi}
\affiliation{Department of Theoretical Physics, University of Szeged, Tisza 
Lajos k\"{o}r\'{u}t 84, H-6720 Szeged, Hungary}

\begin{abstract}
We propose a scheme to implement the one-dimensional coined quantum walk with 
electrons transported through a two-dimensional network of spintronic 
semiconductor quantum rings. The coin degree of freedom is represented by the 
spin of the electron, while the discrete position of the walker corresponds to 
the label of the rings in one of the spatial directions in the network. We 
assume that Rashba-type spin-orbit interaction is present in the rings, the 
strength of which can be tuned by an external electric field. The geometry of 
the device, together with the appropriate spin-orbit interaction strengths, 
ensure the realization of the coin-toss (i.e. spin-flip) and the step operator.  
\end{abstract}


\maketitle

\section{Introduction}

Quantum walks \cite{ADZ93} (QWs) are generalizations of classical random walks 
to quantum systems. For reviews on quantum walks see Refs.~\onlinecite{K03} and 
\onlinecite{Ko2008}. The unitary time evolution of the walk can be either 
discrete \cite{M96a,M96b,W01} leading to coined QWs or continuous. \cite{FG98,CFG02} 
Recently, quantum walks have been shown to be efficient tools 
to design quantum algorithms. \cite{K06a,S08} Coined QWs were applied in the 
first algorithmic proposal \cite{SKW03} for the quantum walk search on a 
hypercube.

Several experimental schemes have been proposed to realize coined QWs including 
ion traps, \cite{TM02} microwave cavities, \cite{SB03} cavity quantum 
electrodynamics, \cite{DHZ04} superconducting quantum electrodynamics, 
\cite{Xu2008} arrays of optical traps, \cite{EM05} ground state atoms 
\cite{DR02} and ultracold Rydberg atoms \cite{CR06} in optical lattices, linear 
optics, \cite{ZD02,HBF03,FH04,KB05,JPK04,PA07} Bose-Einstein condensation, 
\cite{C06} coherent atomic system with electromagnetically induced transparency 
\cite{Li2008} and in a Fabry-Perot cavity. \cite{KRS03} An experimental 
implementation of a continuous time QW on a two-qubit NMR quantum computer 
\cite{DLX03} has already been carried out. In another experiment waveguide 
lattices were employed to realize continuous time quantum walks. 
\cite{PL08} Up to now there is no experimental realization of QWs in 
solid-state systems. Implementation of the continuous time walk has been 
proposed with tunnel coupled quantum dots, \cite{T07} whereas  in the proposal of 
Ref.~\onlinecite{HB09} electrons in  lateral quantum dots would realize the step 
operator of a quantum walk. In another proposal, \cite{MW2008} stimulated Raman 
adiabatic passage  operations are applied to an electron in a quantum dot to 
realize the coined quantum walk on the line.

In this paper we consider a possible scheme for the implementation of a coined 
QW on the line, based on the ballistic transport of an electron through a 
two-dimensional series of semiconductor quantum rings. The spin of the electron 
plays the role of the coin, and its position in one of the spatial directions 
corresponds to the position of the walker along the line. The shift along the 
perpendicular spatial direction can be considered as the discrete time-steps.

Quantum rings, \cite{NMT99} which are the building blocks of our proposal are 
nanoscale rings, fabricated in semiconductor heterostructures, such as InGaAs/
InAlAs \cite{BKSN06,NB07} or HgTe/HgCdTe \cite{KTHS06} where the 
control of the electron spin is possible due to e.g. spin-orbit interaction 
(SOI), and quantum interference. A widely studied type of SOI in such 
heterostructures is the so-called Rashba SOI, \cite{R60} which 
originates from the structural inversion asymmetry of the interface 
confining potential that is accompanied by an electric field directed along the 
normal of the interface, coupling the electron spin and orbital motion.
\cite{FMA07} This type of SOI has gained much 
interest due to its tunability with external gate voltages, 
\cite{NATE97,G00} offering possible applications in semiconductor spin 
electronics, or spintronics. \cite{ALS02} 

Quantum rings with Rashba-type SOI have been shown to have versatile 
applicability. A large variety of single-qubit quantum gates can be realized by 
quantum rings connected with two external leads, \cite{FMBP05} where the spin of 
the electron plays the role of the qubit. Quantum rings with three terminals can 
be used as electron spin beam splitters, i.e., to polarize the spin of the 
electron on the outputs with different spin directions. \cite{FKBP06} 
Two-dimensional arrays of quantum rings \cite{BKSN06,NB07} also show 
nontrivial spin transformations at the outputs of the network. 
\cite{KFBP08b,FKBP08} 

We focus on narrow rings in the ballistic (coherent) regime, 
\cite{KFBP08b,BO08b} where a one-dimensional model provides appropriate
description. We propose a two-dimensional network of such two- and 
three-terminal rings of appropriate size and externally tunable Rashba SOI 
strength for the implementation of the coined QW on the line. We show that with 
appropriately chosen parameters, one can achieve reflectionless operation which 
is necessary for the unitarity of the walk.

In usual experimental situations when ballistic properties are investigated, the 
current is initiated by a potential difference on the two sides of the sample 
with metallic contacts. In order to achieve the highest possible coherence 
length, experiments are carried out at very low temperatures (few hundred mK). 
The conduction is due to electrons with energies very close to the Fermi energy 
of the material, i.e., the problem can be considered a stationary one. The spin 
state of the electrons originating from the metallic contact is generally not a 
pure quantum mechanical state, it is a mixture. However, this means no 
significant restriction, as their spin can be made polarized by eg. a 
three-terminal quantum ring. \cite{FKBP06}

The paper is organized as follows. In Sec.~\ref{QWsec} we give a short overview 
of the model of the coined QW on the line. In Sec.~\ref{unitsec} we present the 
functional unit of the scheme: we start with the model we use in 
Sec.~\ref{modelsec}, then in Sec.~\ref{Hsec}, we show the ring that performs the 
coin-toss, and then, in Sec.~\ref{stepsec}, the ring, which is responsible for 
the step operation. In Sec.~\ref{intsec} we show, how a three-terminal ring can 
be used to ensure interference at intermediary positions in the network. In 
Sec.~\ref{schemesec}, we present the proposed scheme to implement the coined QW 
with quantum rings. Finally, we summarize our results in Sec.~\ref{sumsec}.

\section{The coined quantum walk on the line}
\label{QWsec}

In the classical random walk on the line, the walker tosses a coin before each 
step. The direction of the step is determined by the actual state of the coin, 
i.e., the walker takes a step to the left if the coin is heads or to the right 
if the coin is tails (or vice versa). The quantum analog of such a walk uses a 
quantum coin, the state of which can be a linear combination of the classical 
heads and tails, or mathematically, any state of a 'coin' Hilbert space 
$\mathcal{H}_{C}$, spanned by the two basis states $\left\lbrace\left| L\right>,
\left| R \right> \right\rbrace$, where $L$ ($R$) stand for 'left' ('right'). The 
positions of the walker also span a Hilbert space $\mathcal{H}_{P}=
\left\lbrace \left| i \right> : i \in \mathbf{Z} \right\rbrace $ with 
$\left| i \right>$ corresponding to the walker localized in position $i$. The 
states of the total system are in the space $\mathcal{H}=\mathcal{H}_{C}\otimes
\mathcal{H}_{P}$. The conditional step of the walker dependent on the state of 
the coin, can be described by the unitary operation
\begin{equation}
 S = \left| L \right> \left< L \right| \otimes \sum_{i} \left| i+1 \right>
 \left< i \right| + \left| R \right> \left< R \right| \otimes \sum_{i} 
 \left| i-1 \right> \left< i \right|. \label{S}
\end{equation}
The coin-toss is realized by a unitary operation $C$ acting in the space 
$\mathcal{H}_{C}$. The QW of $N$ steps is defined as the transformation $U^{N}$, 
where $U$, acting on $\mathcal{H}=\mathcal{H}_{C} \otimes \mathcal{H}_{P}$ is 
given by
\begin{equation}
 U = S \cdot \left( C \otimes I \right), \label{U}
\end{equation}
with $I$ being the identity operator. A frequently used balanced unitary coin is 
the Hadamard coin $H$, which is represented by a matrix in which each element is 
of equal magnitude.

In the QW the coin state is not measured during intermediate iterations, thus 
quantum correlations between different positions are kept, leading to 
interference in subsequent steps. We note that this interference causes a 
radically different behavior from that of the classical random walk. In 
particular, the probability distribution of the walk on the line does not 
approach a Gaussian -- it leads to a double-peaked distribution -- and the 
variance $\sigma^{2}$ is not linear in the number of steps $N$, it scales with 
$\sigma^{2} \sim N^{2}$, which implies that the expected distance from the 
origin is of order $\sigma \sim N$, i.e. the quantum walk propagates 
quadratically faster than the classical random walk. This property is at the 
heart of algorithmic applications.

In our proposal, the walker is the electron, which is transported through a 
two-dimensional network of quantum rings. The coin states $\left\lbrace \left| L
\right>, \left| R \right> \right\rbrace$ are represented by appropriate 
orthogonal states of the electron spin, while the Hilbert-space 
$\mathcal{H}_{P}$ is characterized by the discrete positions of the electron in 
one spatial direction in the network. In the following sections we will show, 
that quantum rings with appropriate radius and SOI strength act essentially as 
the unitary transformations $H$ and $S$. Namely, a ring connected with two leads 
acts essentially as the Hadamard operation $H$, while a totally symmetric 
three-terminal ring can implement the step operation $S$ given by Eq. (\ref{S}). 
These rings together (which we will call a functional unit) act as the unitary 
transformation $U$, given by Eq. (\ref{U}). We note that this kind of operation 
of the network is based on the fact that practically zero reflection can be 
ensured at each individual ring, by appropriately choosing the strength of the 
Rashba SOI and the geometry. If considerable reflections were present in the 
network, the state of the walker would spread out in two-dimensions, and the 
analogy with the model of the QW could not be made. 

\bigskip

\section{The functional unit of the scheme}
\label{unitsec}

In this section we propose two- and three-terminal rings to be buliding blocks 
of the QW scheme, and introduce the unit, which implements a single step of the 
QW with a Hadamard coin. It consists of a two-terminal ring realizing the 
Hadamard transformation (Hadamard-ring) and a subsequent three-terminal ring 
which performs the step operation (step-ring).

\subsection{The model of quantum rings}
\label{modelsec}

We consider a narrow ring of radius $a$ situated in the $x-y$ plane. The 
Hamiltonian in single-electron picture, in the presence of Rashba SOI is given 
by \cite{MPV04,MMK02}
\begin{equation}
H\!=\!\hbar \Omega \!\left[ \!\left( \!-i\frac{\partial }{\partial \varphi }%
\!+\!\frac{\omega }{2\Omega }(\sigma _{x}\cos \varphi \!+\!\sigma _{y}\sin
\varphi )\!\right) ^{2}\!-\!\frac{\omega ^{2}}{4\Omega ^{2}}\!\right] ,
\label{Ham}
\end{equation}
where $\varphi $ is the azimuthal angle of a point on the ring, 
$\hbar \Omega =\hbar ^{2}/2m^{\ast }a^{2}$ is the dimensionless kinetic energy, 
with $m^{\ast }$ denoting the effective mass of the electron, and 
$\omega =\alpha/\hbar a$ is the frequency associated with the SOI, which can be 
changed by an external gate voltage that tunes the value of $\alpha$. 
\cite{NATE97} The energy eigenvalues and the corresponding eigenstates of this 
Hamiltonian can be calculated analytically.\cite{MPV04,FMBP05} For a ring with 
leads attached to it, the spectrum is continuous; all positive energies can 
appear, and they are 4-fold degenerate. This degeneracy is related to (i) two 
possible eigenspinor orientations and to (ii) the two possible (clockwise and 
anticlockwise) directions in which currents can flow. The state of the incoming 
electron is considered to be a plane wave with wavenumber $k$. By energy 
conservation, its energy (given by $E=\hbar^{2}k^{2}/2m^{\ast}$) determines the 
solutions in the rings. At the incoming lead - ring, and the outgoing leads - 
ring junctions (see Fig.~\ref{3leadringfig}), Griffith's boundary conditions 
\cite{G53} are applied; that is, the net spin current density at a certain 
junction has to vanish, and we also require the continuity of the spinor valued 
wave functions. (We note that there are other frequently used boundary 
conditions as well, \cite{BIA84,V09} they are usually based on detailed physical 
description of the junctions, e.g. the non-ideality of the couplings.) The 
solution of the scattering problem in two- and three-terminal rings with one 
input has been investigated in Refs. \onlinecite{MPV04,FMBP05,FKBP06,KFBP08} and 
for a general boundary condition in Ref. \onlinecite{KFBP08b}. For the sake of 
completeness, these results are summarized in the Appendix.

\subsection{The Hadamard ring}
\label{Hsec} 

In this section we consider the quantum ring, which implements the Hadamard 
coin-toss $H$. As it has been shown in Ref.~\onlinecite{FMBP05}, a two-terminal 
ring acts as a linear transformation on the spin state of the electron (see Eq. 
(\ref{2leadtrans}) of the Appendix). When the parameter $\omega/\Omega$ 
characterizing the strength of the Rashba SOI is equal to $-1$ in a ring in 
which the two terminals are in a diametrical position (see 
Fig.~\ref{3leadringfig} without lead \textbf{2}, and $\gamma1=\pi$) and there is 
only one input (i.e. $f_{1}=0$), the spin state $r_{1}$ of the transmitted 
electron is essentially the Hadamard transform of the incoming spinor, i.e. the 
transmission matrix corresponding to the ring is given by \cite{FMBP05}
\begin{equation}
\hat{T} = c \frac{1}{\sqrt{2}}
\left( \begin{matrix}
       1 & 1 \\
       -1 & 1
       \end{matrix} \right),
\end{equation}
where
\begin{equation}
c = \frac{8ikaq}{\hat{y}}
\sin\left(q\pi\right)\sin\left(w\frac{\pi}{2}\right), \notag
\end{equation}
and
\begin{eqnarray}
\hat{y}&=& k^{2}a^{2}\left[1-
\cos\left(2q\pi\right)\right] + 4ikaq \sin\left(2q\pi\right) \notag \\
   && -4q^{2} \left[\cos\left(w\pi \right)+\cos\left(2q\pi\right)\right], \notag
\end{eqnarray}
with $w=\sqrt{1+\left(\omega/\Omega\right)^{2}}$ and 
$q=\sqrt{\left(\omega/2\Omega\right)^{2}+k^{2}a^{2}}$.

In the most general case the transmission efficiency of the quantum ring is less 
than 1, i.e. there is a nonzero probability for the electron to be reflected 
into the terminal through which it enters the ring. In order for the Hadamard 
ring to operate in a unitary way, the transmission probability 
$\left|c\right|^{2}$ has to be equal to unity, which can be given by the 
following condition:
\begin{equation}
k^{2}a^{2}\sin^{2}\left(q\pi\right) = -2q^{2} \left[ \cos\left(w\pi\right) +
\cos \left(2q\pi\right)\right].
\end{equation}
This condition can be satisfied for an appropriate radius of the ring as can be 
seen in Fig. 2 of Ref.~\onlinecite{FMBP05}. We note that the wave number $k$ of 
the electron is determined by the Fermi level of the semiconducting material in 
which the quantum ring is fabricated. For InGaAs the Fermi energy is 11.13 meV, 
corresponding to $k_{F}a=20.4$ for a ring of radius 0.25 $\mu$m. For the sake of 
definiteness we are going to focus on this material.

\subsection{The step ring}
\label{stepsec}

\begin{figure}[tbh]
\includegraphics[width=5.5cm]{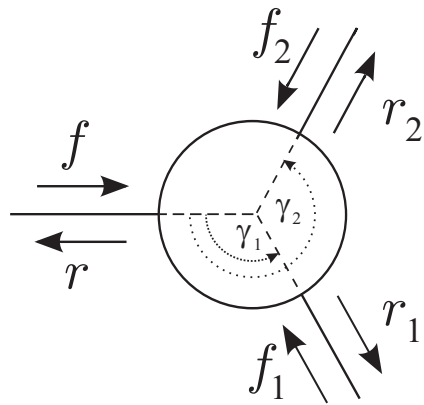}
\caption{Three-terminal ring with the most general boundary condition. $f$, $r$, 
etc. denote two-component spinors, the arrows indicate the direction of the 
corresponding wave number.}
\label{3leadringfig}
\end{figure}

For the step operation to be implemented we use a three-terminal ring which has 
only one input lead and two output leads, and the leads are equally separated 
from each other (i.e. $\gamma_{1}=2\pi/3$, $\gamma_{2}=4\pi/3$, $f_{1}=0$ and 
$f_{2}=0$ in Fig. \ref{3leadringfig}). The outgoing spinors $r_{1}$ and $r_{2}$ 
are linear transforms of the incoming spinor $f$, the transformations being 
given by Eqs. (\ref{T_3term}) and (\ref{T2_3term}), respectively. 

In the following we will recall the previously obtained result, 
\cite{FKBP06,KFBP08} that a totally symmetric ring, which is shown in 
Fig.~\ref{3leadringfig}, can be considered an electron spin polarizer (the
derivation of this property is summarized in the Appendix). In other
words, there are two orthogonal input spin states, for one of which, there is no 
output in lead \textbf{1}, while for the other, there is no output in lead 
\textbf{2}. We will show, that we can take advantage of this property 
if we choose to define the coin states to be these states, and thus obtain a 
ring which performs the step operation.

%%%%%%%%%%%%%%%%%%%%%%%%%%%%%%%%%%%%%%%%%%%%%%%%%%%%%%%%%%%%%%%%%%%%%%%%%%%%%%%%
%%%%%%%%%%%%%%%%%%%%%%%%%%%%%%%%%%%%%%%%%%%%%%%%%%%%%%%%%%%%%%%%%%%%%%%%%%%%%%%%

As derived in Refs.~\onlinecite{FKBP06}, \onlinecite{KFBP08}, and in the 
Appendix, if the equations 
\begin{subequations} \label{pol_cond}
\begin{eqnarray}
\cos \left(w\pi \right) &=& 2 \cos \left(q \frac{2\pi}{3}\right),
\label{pol_cond_1} \\
\sin \left(w\pi \right) &=& \frac{ka}{q}\sin\left(q \frac{2\pi}{3}\right), 
\label{pol_cond_2}
\end{eqnarray}
\end{subequations}
are satisfied simultaneously then the ring polarizes a totally unpolarized 
input, given by the density matrix $\varrho$, proportional to the identitiy. 
The polarized spinors exiting at the two outputs 
\begin{equation}
\left|\phi_{1}\right> = \begin{pmatrix}
                        -e^{-i\frac{\pi}{3}} \sin \frac{\theta}{2} \\
                         e^{i\frac{\pi}{3}} \cos \frac{\theta}{2}
                        \end{pmatrix},
\quad
\left|\phi_{2}\right> = \begin{pmatrix}
                        e^{-i\frac{2\pi}{3}} \cos \frac{\theta}{2} \\
                        e^{i\frac{2\pi}{3}} \sin \frac{\theta}{2}
                        \end{pmatrix}, \label{pol_output}
\end{equation}
are the eigenstates with nonzero eigenvalues $\eta_{n}$ of the output density 
matrices $\varrho_{n}=\frac{1}{2}\tilde{T}_{n}(\tilde{T}_{n})^{\dagger}$ 
($n=1,2$), where $\tilde{T}_{n}$ are given by Eqs. (\ref{T12_3term_p}) of the 
Appendix. The corresponding eigenvalues, which describe the transmission 
probability in the outputs are
\begin{equation}
\eta_{1}=\eta_{2}=\frac{128k^{4}a^{4}q^{2}\sin^{2}\left(q\frac{2\pi}{3}\right)}
{\left|\tilde{y}\right|^{2}}, \label{eta}
\end{equation}
where 
\begin{eqnarray}
\tilde{y} &=& 8q^{3}\left[\cos\left(w\pi\right)+\cos\left(2q\pi\right)\right]
-12ikaq^{2}\sin\left(2q\pi\right) \notag \\
&& +6k^{2}a^{2}q \left[ \cos\left(2q\pi\right)
 - \cos\left(q \frac{2\pi}{3}\right)\right] \notag \\
&& +ik^{3}a^{3}\left[ 3\sin\left(q \frac{2\pi}{3}\right)
-\sin\left(2q\pi\right)\right].
\end{eqnarray}
If we determine the spinors $\left|\phi_{n}^{0}\right>$ 
($n=1,2$) annuled by the transmission matrices 
$\tilde{T}_{n}\left|\phi_{n}^{0}\right>=0$
\begin{equation}
\left|\phi_{1}^{0}\right> = \begin{pmatrix}
                            \cos \frac{\theta}{2} \\
                            \sin \frac{\theta}{2}
                            \end{pmatrix},
\quad
\left|\phi_{2}^{0}\right> = \begin{pmatrix}
                            -\sin \frac{\theta}{2} \\
                            \cos \frac{\theta}{2}
                            \end{pmatrix}, \label{pol_zero}
\end{equation}
then it can easily be seen, that if the input state is the 
$\left|\phi_{1}^{0}\right>$ ($\left|\phi_{2}^{0}\right>$) pure state, then the 
transmission into output lead \textbf{1} (\textbf{2}) will be zero, while the 
spin direction of the output in lead \textbf{2} (\textbf{1}) will be given by 
$\left|\phi_{2}\right>$ ($\left|\phi_{1}\right>$), i.e.:
\begin{eqnarray}
\left|\phi_{1}^{0}\right> &\rightarrow &\left\{
\begin{array}{ll} 
\left|\phi_{2}\right> \text{in lead \textbf{2}}\\
\text{no output in lead \textbf{1}} &
\end{array}%
\right.   \notag \\
\left|\phi_{2}^{0}\right> &\rightarrow &\left\{
\begin{array}{ll}
\text{no output in lead \textbf{2}} & \\
\left|\phi_{1}\right> \text{in lead \textbf{1}} & 
\end{array}%
\right. 
\end{eqnarray}

These --orthogonal-- input states are suitable to represent the coin states 
$\left\lbrace \left|L \right>, \left| R \right> \right\rbrace$ in the QW as they 
form a basis in the two-dimensional space of the electron spin, and the 
polarizing three-terminal ring acts on them as the step operator in the QW: if 
the input spin (coin) state is $\left|\phi_{1}^{0}\right>$
($\left|\phi_{2}^{0}\right>$) the electron is transmitted into the output lead 
\textbf{2} (\textbf{1}), i.e. the walker 'takes a step to the left (right)'. The 
change in the spin direction at the outputs given by Eq. (\ref{pol_output}) 
means that the states $\left|\phi_{1}\right>$ and $\left|\phi_{2}\right>$ are 
rotated versions of the two orthogonal inputs $\left|\phi_{2}^{0}\right>$ and 
$\left|\phi_{1}^{0}\right>$, respectively, where the rotation is around the 
z-axis by the angle of the given output lead. As we will see in the following, 
these rotations can be reversed by the application of appropriate rings.

\begin{figure}[tbh]
\begin{center}
\includegraphics*[width=8.0cm]{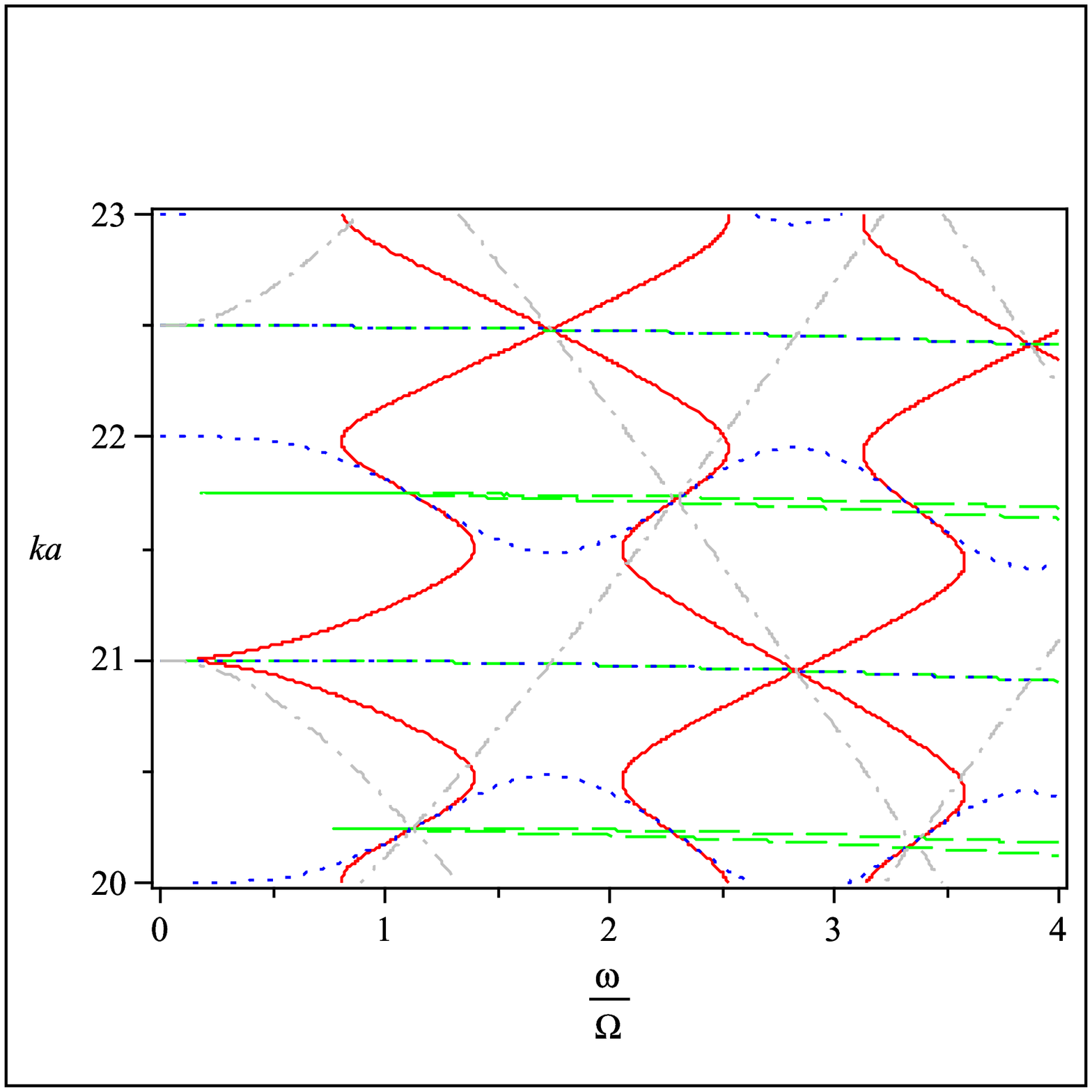} \vspace{%
-0.4cm}
\end{center}
\caption{(Color online) Determination of the parameter values corresponding to 
the step ring: Eqs. (\ref{pol_cond_1}) and (\ref{pol_cond_2}) are satisfied 
along the blue (dotted) and grey (dashdotted) lines, respectively, while Eqs. 
(\ref{step_0refl_a}) and (\ref{step_0refl_b}) are satisfied along the red 
(solid) and green (dashed) lines, respectively. At the intersection of these 
four curves the totally symmetric ($\gamma_{1}=2\pi/3$, $\gamma_{2}=4\pi/3$) 
three-terminal ring behaves as a perfect polarizing device with practically zero 
reflection, i.e. as the step ring.}
\label{paramfig}
\end{figure}

In order for the transformation to be unitary the step-ring also has to be 
reflectionless, that is the transmission probabilities into the two outputs 
given by Eq. (\ref{eta}) should be equal to unity, i.e. $\eta_{1}=\eta_{2}=1/2$. 
It can be easily verified that this condition can be formulated by the following
equations
\begin{subequations} \label{step_0refl}
\begin{equation}
k^{2}\!a^{2}\!\sin^{2}\!\left(\!q\frac{2\pi}{3}\!\right)\!\cos\left(
\!q\frac{2\pi}{3}\!\right)
\!+\!q^{2}\!\left[\cos\left(w\pi\right)\!+\!\cos\left(2q\pi\right)\right]=0, 
\label{step_0refl_a} 
\end{equation}
\begin{equation}
k^{2}\!a^{2}\!\sin^{3}\!\left(\!q\frac{2\pi}{3}\!\right)
\!+\!q^{2}\!\sin\left(2q\pi\right)=0. \label{step_0refl_b}
\end{equation}
\end{subequations}
which, for an appropriate combination of the parameters 
$\left\lbrace a,\omega/\Omega,\gamma_{2}\right\rbrace$ can be satisfied together 
with Eqs. (\ref{pol_cond_1}) and (\ref{pol_cond_2}) as can be seen in 
Fig.~\ref{paramfig} for the experimentally feasible range of the parameters.

%%%%%%%%%%%%%%%%%%%%%%%%%%%%%%%%%%%%%%%%%%%%%%%%%%%%%%%%%%%%%%%%%%%%%%%%%%%%%%%%
%%%%%%%%%%%%%%%%%%%%%%%%%%%%%%%%%%%%%%%%%%%%%%%%%%%%%%%%%%%%%%%%%%%%%%%%%%%%%%%%

In order to use the same building blocks (i.e. the Hadamard ring and the step 
ring again) for later steps, the rotations on the basis states introduced by the 
step ring need to be removed. This can be done eg. by the application of two 
two-terminal rings which act as $\hat{T}^{\left(1\right)}=U_{2\pi/3}^{-1}$ and 
$\hat{T}^{\left(2\right)}=U_{4\pi/3}^{-1}$, where 
\begin{equation}
U_{\gamma}=\left( \begin{matrix}
                  e^{-i\frac{\gamma}{2}} & 0 \\
                  0 & e^{i\frac{\gamma}{2}}
                  \end{matrix} \right). \label{U_gamma}
\end{equation}
Although these conditions can be fullfilled if $\gamma=4\pi/3$, and 
$\gamma=2\pi/3$, respectively, the radius of the rings cannot by made equal to 
that of the step ring, which does not permit the simple attachment of successive 
building blocks. We will show in the following section, that three-terminal 
rings of the same size as the step ring with an appropriate SOI strength, can 
also rotate the spin states in the desired way, as well as allow of the 
continuation of the units.

\section{Interference at intermediary positions}
\label{intsec}

Clearly, the functional units have to be combined so that the walker can arrive 
in any intermediate point on the 'line of the walk' from two directions, i.e., 
interference phenomena can take place. In order to implement this crucial 
property of the QW, we use another quantum ring, which is capable of adding the 
two probabilty amplitudes that both represent the walker at the given point on 
the 'line of the walk', as well as rotating the spins back into the basis states 
$\left| \phi_{1}^{0}\right>$ and $\left| \phi_{2}^{0}\right>$. Now we show, that 
this can be done with a completely symmetric three-terminal ring which has the 
same radius as the step ring, and in which the magnitude of the SOI strength is 
the same, but its direction is opposite.

If two leads of a symmetric, three-terminal ring are considered as inputs and 
the other terminal as an output (see Fig.~\ref{3leadringfig} with $f=0$), the 
matrices of the one-input case, given by Eqs. (\ref{T_3term}) and 
(\ref{T2_3term}), are enough to handle the problem. \cite{KFBP08b} Namely, we 
can consider the two inputs $f_{i}$ ($i=1,2$) separately, and determine the 
corresponding matrices. The outputs in each terminal in the superposed problem 
will consist of contributions from both inputs. Considering $f_{1}$ ($f_{2}$) as 
the only input, the transmission matrices in the reference frame of $f_{1}$ 
($f_{2}$) are the same as those for the input $f$, given by Eqs. (\ref{T_3term}) 
and (\ref{T2_3term}). In order to get the contributions to the output spinors 
($r,r_{1},r_{2}$) for the input $f_{1}$ ($f_{2}$) in the reference 
frame of $r$, the matrices need to be rotated by the angle of 
$\gamma_{1}=2\pi/3$ ($\gamma _{2}=4\pi/3$). Furthermore, since we 
have considered a propagation of the electron from the left to the right, the 
symmetric three-terminal ring we want to use to add the two (spin-dependent) 
probability amplitudes has to be rotated by an angle of $\pi$ with respect to 
Fig.~\ref{3leadringfig}. This means an additional rotation of each 
matrix by $\pi$.

%%%%%%%%%%%%%%%%%%%%%%%%%%%%%%%%%%%%%%%%%%%%%%%%%%%%%%%%%%%%%%%%%%%%%%%%%%%%%%%%
\begin{figure}[tbh]
\begin{center}
\includegraphics*[width=5.0cm]{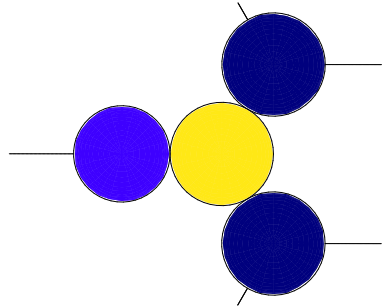} \vspace{%
-0.4cm}
\end{center}
\caption{(Color online) The geometry of the functional unit of the scheme, where 
the rotations introduced by the step ring are removed by symmetric 
three-terminal rings of the same size as the step ring, but with opposite SOI 
strength. The colors indicate the value of the SOI strength ($\omega/\Omega$): 
light blue, yellow and dark blue corresponding to -1, 2.27 and -2.27, 
respectively. The radius of the Hadamard ring is $a_{H}=0.248$ $\mu$m, while 
that of the other rings is $a_{S}=0.266$ $\mu$m.}
\label{schemefig2}
\end{figure}
%%%%%%%%%%%%%%%%%%%%%%%%%%%%%%%%%%%%%%%%%%%%%%%%%%%%%%%%%%%%%%%%%%%%%%%%%%%%%%

If the radius of the above mentioned ring is the same as that of the step-ring,
and the applied SOI strength ($\omega/\Omega$) is of the same magnitude, but 
opposite direction (in which case the polarization condition given by 
Eq. (\ref{pol_cond}), and the condition for zero reflection of the input, given 
by Eq. (\ref{step_0refl}) also hold), then by using Eqs. (\ref{T12_3term_p}) of 
the Appendix, it can easily be shown that zero reflection in the two input 
arms without any transmission from one input lead into the other (i.e. 
$r_{1}=r_{2}=0$) is automatically guaranteed. Additionally, the probability of 
transmission from the two inputs into the output is the same, and the coin 
states $\left|\phi_{1}\right>$ and $\left|\phi_{2}\right>$ are rotated into 
$\left|\phi_{2}^{0}\right>$ and $\left|\phi_{1}^{0}\right>$, respectively. 
Hence, such a  ring will be able to transform the two inputs into the 
superposition of the basis states ($\left|\phi_{1}^{0}\right>$ and 
$\left|\phi_{2}^{0}\right>$) with the same weights.

This ring can also be used for the same purpose as the two-terminal rings 
mentioned in the previous section. Fig.~\ref{schemefig2} shows the functional 
unit of the scheme. The colors of the rings denote the value of the SOI 
strength, which together with the appropriate radius of the ring, guarantee that 
no reflection occurs at the inputs. The advantage of using this symmetric 
three-terminal ring is that it has the same size as the step ring, providing a 
more symmetrical arrangement for the QW (see Fig.~\ref{networkfig}). 
Additionally, measuring currents at the junctions indicated by the short lines 
in Fig.~\ref{schemefig2}, can be used for determining the functionality of the 
device: No currents leave the network through these leads under ideal 
circumstances.

\section{The proposed scheme} \label{schemesec}

%%%%%%%%%%%%%%%%%%%%%%%%%%%%%%%%%%%%%%%%%%%%%%%%%%%%%%%%%%%%%%%%%%%%%%%%%%%%%%
\begin{figure}[tbh]
\includegraphics*[width=8.0cm]{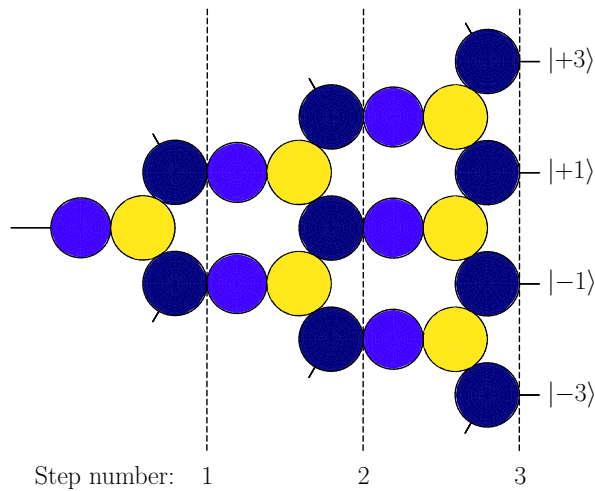}
\caption{(Color online) The geometry of the device for three steps. The colors 
and the radii of the rings are the same as in Fig.~\ref{schemefig2}. The 
vertical dashed lines indicate where the discrete steps take place along the 
horizontal direction. The vertical direction corresponds to the 'line of the 
walk', along which the walk takes place. We have indicated in the vertical 
direction on the right hand side of the figure the discrete position states of 
the walker (the electron) on the line after three steps.}
\label{networkfig}
\end{figure}
%%%%%%%%%%%%%%%%%%%%%%%%%%%%%%%%%%%%%%%%%%%%%%%%%%%%%%%%%%%%%%%%%%%%%%%%%%%%%%

Our scheme uses several functional units as building blocks for the 
implementation of the QW on the line, thus actually it corresponds to a 
two-dimensional displacement of the walker (the electron). One spatial dimension 
represents the 'line of the walk' along which the walk is realized, while the 
role of the other dimension is twofold. First it is necessary from the technical 
point of view, it is needed for the transformations (coin-toss, and step) to be 
made, but it is also related to the discrete time steps of the walk: the number 
of the functional units increases in this direction, according to the possible 
positions of the walker that completes increasing number of steps. In other 
words, the notion of time enters the otherwise time-independent scheme via this 
spatial direction.

Fig. \ref{networkfig} shows a 
device capable of implementing three steps of a QW on the line with a Hadamard 
coin. The colors denote the value of the SOI strength in the rings, which 
together with the appropriate radius of the ring, guarantee that no reflection 
occurs at the inputs. For the removal of the rotations of the spins the same 
symmetric three-terminal ring is used (see Fig.~\ref{schemefig2}) as the one 
which adds the probability amplitudes at intermediary positions. The vertical 
dashed lines indicate where the discrete steps take place along the horizontal 
direction. The vertical direction corresponds to the 'line of the walk', along 
which the walk takes place. On the right hand side of the figure we have 
indicated the discrete position states of the walker (the electron) on the 'line 
of the walk' after three steps. As the transmission probabilities are 
proportional to the value of the current, by measuring the currents on these 
terminals the distribution characteristic of a QW appears.

\section{Conclusion}
\label{sumsec}

We have proposed a scheme for the implementation of the coined QW on the line, 
where the coin is the spin of the electron, quantum rings are used to realize
the coin-toss and the step operations, and the shift of the electron in one 
spatial direction corresponds to the walk along the line.

Let us note that our scheme is based on a one-dimensional model of 
quantum rings, that assumes single channel ballistic transport. Although spin 
coherence lengths of 100 and 350 $\mu$m have been found in bulk GaAs \cite{KA99} 
and Si \cite{HMA07} samples, respectively, the coherence lengths of the orbital 
wave function are typically two magnitudes shorter, even in modulation doped 
heterostructures, where the mobility is higher. In the case of InGaAs/InAlAs, 
the coherence lengths of the orbital wave function are typically in the range of 
a few microns, \cite{BKSN06} which means a severe constrain for our QW scheme. 
On the other hand, there are samples, where transport is due to many channels 
in the ring, \cite{KTHS06} for which our results are not directly applicable. 
It is beyond the scope of this paper to investigate in detail how phase 
destroying events affect the functionality of the proposed network, but 
preliminary results indicate that the functionality can tolerate moderate level 
of scattering induced errors, thus a few steps of the QW could be implemented.

Our aim was to demonstrate the possibility of a scheme for the QW with 
semiconductor quantum rings. Further optimization on the number of rings, and 
the geometry of the network might be possible.

\section*{Acknowledgement}
This work was supported by the Hungarian Scientific Research Fund (OTKA) under 
Contracts No. T48888, and No. T49234. P.F. was supported by a 
J. Bolyai grant of the Hungarian Academy of Sciences. We thank M. G. Benedict 
for helpful discussions.

\section*{Appendix}

For the sake of completeness we present the analytic expressions for the 
transmission matrices of one-input two- and three-terminal rings, in which 
Rashba SOI is present. These matrices are obtained by applying Griffith's 
boundary conditions \cite{G53} at the junctions between the incoming lead and 
the ring, and the outgoing lead(s) and the ring (see Fig.~\ref{3leadringfig}), 
that is, requiring vanishing net spin current density at a certain junction, as 
well as the continuity of the spinor valued wave functions.

The transmission matrix of a two-terminal ring (see Fig.~\ref{3leadringfig} 
without lead \textbf{2} and $\gamma_{1}=\gamma$) is given by \cite{FMBP05}
\begin{eqnarray}
\hat{T}_{\uparrow \uparrow} &=& \frac{4ikaq}{\hat{y}}e^{-i\frac{\gamma}{2}}
\left( h \cos^{2} \frac{\theta}{2} + h^{\ast} \sin^{2} \frac{\theta}{2} \right), 
\notag \\
\hat{T}_{\uparrow \downarrow} &=& \frac{4ikaq}{\hat{y}}e^{-i\frac{\gamma}{2}}
\sin \frac{\theta}{2} \cos \frac{\theta}{2} \left( h - h^{\ast} \right), 
\notag \\
\hat{T}_{\downarrow \uparrow} &=& e^{i \gamma} \hat{T}_{\uparrow \downarrow}, 
\label{2leadtrans} \\
\hat{T}_{\downarrow \downarrow} &=& \frac{4ikaq}{\hat{y}}e^{i\frac{\gamma}{2}}
\left( h \sin^{2} \frac{\theta}{2} + h^{\ast} \cos^{2} \frac{\theta}{2} \right), 
\notag
\end{eqnarray}
where
\begin{eqnarray}
h &=& e^{i\frac{w}{2}\gamma}\left[\sin\left(q\left(2\pi-\gamma\right)\right)
      -e^{-iw\pi} \sin \left(q\gamma\right) \right], \notag \\
\hat{y}&=& k^{2}\!a^{2}\!\left[\cos\left(2q\left(\pi\! -\!\gamma\right)\right)
      \!-\!\cos \left( 2q\pi \right) \right]\!+\!4ikaq \sin \left( 2q\pi \right) 
\notag \\
  &&  -4q^{2}\left[\cos\left(w\pi\right)+\cos\left(2q\pi\right)\right], 
\notag \\
\theta&=&\arctan \left(-\omega/\Omega\right) \notag
\end{eqnarray}
with $w=\sqrt{1+\omega^{2}/\Omega^{2}}$, and 
$q=\sqrt{\left(\omega/\Omega\right)^{2}+k^{2}a^{2}}$.

The transmission matrices of a totally symmetric (i.e. $\gamma_{1}=2\pi/3$ and 
$\gamma_{2}=4\pi/3$) three-terminal ring, which is shown in 
Fig.~\ref{3leadringfig} (with $f_{1},f_{2}=0$) are given by \cite{FKBP06,KFBP08}
\begin{eqnarray}
(\tilde{T}_{1})_{\uparrow \uparrow}\!&=&\!
\frac{8kaq}{\tilde{y}} e^{i\frac{2\pi}{3}}\!
\left[\cos^{2}\!\frac{\theta}{2}\!\left(h_{1}+h_{2}\right)
\!+\!\sin^{2}\!\frac{\theta}{2}\!\left(h_{1}^{\ast}-h_{2}^{\ast}\right)\right], 
\notag \\
(\tilde{T}_{1})_{\uparrow \downarrow}\!&=&\!
\frac{8kaq}{\tilde{y}} 
e^{i\frac{2\pi}{3}}\!\sin\!\frac{\theta}{2}\cos\!\frac{\theta}{2}
\left[\left(h_{1}+h_{2}\right)\!-\!\left(h_{1}^{\ast}-h_{2}^{\ast}\right)\right], 
\notag \\
(\tilde{T}_{1})_{\downarrow \uparrow}\!&=&\!
e^{-i\frac{4\pi}{3}} (\tilde{T}_{1})_{\uparrow \downarrow}, \label{T_3term} \\
(\tilde{T}_{1})_{\downarrow \downarrow}\!&=&\!
\frac{8kaq}{\tilde{y}} e^{-i\frac{2\pi}{3}}\!\left[\sin^{2}\!\frac{\theta}{2}
\!\left(h_{1}+h_{2}\right)
\!+\!\cos^{2}\!\frac{\theta}{2}\!\left(h_{1}^{\ast}-h_{2}^{\ast}\right)\right], 
\notag
\end{eqnarray}
\begin{eqnarray}
(\tilde{T}_{2})_{\uparrow \uparrow}&=&(\tilde{T}_{1})_{\downarrow \downarrow}, 
\notag \\
(\tilde{T}_{2})_{\uparrow \downarrow}&=&-(\tilde{T}_{1})_{\downarrow \uparrow}, 
\notag \\
(\tilde{T}_{2})_{\downarrow \uparrow}&=&-(\tilde{T}_{1})_{\uparrow \downarrow}, 
\label{T2_3term} \\
(\tilde{T}_{2})_{\downarrow \downarrow}&=&(\tilde{T}_{1})_{\uparrow \uparrow}, 
\notag
\end{eqnarray}
where
\begin{eqnarray}
h_{1} &=& kae^{iw\frac{\pi}{3}} \sin^{2}\left(q \frac{2\pi}{3}\right), 
\label{h1} \\
h_{2} &=& -iq \left[ e^{iw\frac{\pi}{3}}\sin\!\left(q \frac{4\pi}{3}\right)\!-
\!e^{-iw\frac{2\pi}{3}}
\sin\!\left(q \frac{2\pi}{3}\right) \right], \label{h2} \\
\tilde{y} &=& 8q^{3}\left[\cos\left(w\pi\right)+\cos\left(2q\pi\right)\right]
-12ikaq^{2}\sin\left(2q\pi\right) \notag \\
&& +6k^{2}a^{2}q \left[ \cos\left(2q\pi\right)
 - \cos\left(q \frac{2\pi}{3}\right)\right] \notag \\
&& +ik^{3}a^{3}\left[ 3\sin\left(q \frac{2\pi}{3}\right)
-\sin\left(2q\pi\right)\right]. \label{y_tilde}
\end{eqnarray}

In order for such a ring to polarize a totally unpolarized input that is
described by the density matrix $\varrho$ proportional to the identitiy, the 
output density operators 
$\varrho_{n}=\tilde{T}_{n}\varrho(\tilde{T}_{n})^{\dagger}$ ($n=1,2$) need to be 
projectors
\begin{equation}
\varrho_{n}=\frac{1}{2}\tilde{T}_{n}(\tilde{T}_{n})^{\dagger}=
\eta_{n}\left|\phi_{n}\right>\left<\phi_{n}\right|, \label{varrho}
\end{equation}
where the nonnegative numbers $\eta_{n}$ measure the efficiency of the 
polarizing device. Equation (\ref{varrho}) is equivalent to requiring the 
determinants of $\tilde{T}_{n}(\tilde{T}_{n})^{\dagger}$ to vanish. These 
determinants are equal, and zero if $h_{1}\pm h_{2}=0$, which, using Eqs. 
(\ref{h1}) and (\ref{h2}) can be formulated as
\begin{eqnarray}
\cos \left(w\pi \right) &=& 2 \cos \left(q \frac{2\pi}{3}\right), \\
\sin \left(w\pi \right) &=& \pm\frac{ka}{q}\sin\left(q \frac{2\pi}{3}\right).
\end{eqnarray}
If we focus on the case when condition $h_{1}+h_{2}=0$ holds, then the 
transmission matrices have the simple form
\begin{subequations}
\label{T12_3term_p}
\begin{eqnarray} 
\tilde{T}_{1}&=&\tilde{c}
\left(\begin{array}{cc}
e^{i\frac{2\pi}{3}}\cos^{2}\frac{\theta}{2} & 
e^{i\frac{2\pi}{3}}\sin\frac{\theta}{2}\cos\frac{\theta}{2} \\
e^{-i\frac{2\pi}{3}}\sin\frac{\theta}{2}\cos\frac{\theta}{2} &
e^{-i\frac{2\pi}{3}}\sin^{2}\frac{\theta}{2}
\end{array}\right), 
\label{T_3term_p}  \quad \\
\tilde{T}_{2}&=&\tilde{c}
\left(\begin{array}{cc}
e^{-i\frac{2\pi}{3}}\sin^{2}\frac{\theta}{2} & 
-e^{-i\frac{2\pi}{3}}\sin\frac{\theta}{2}\cos\frac{\theta}{2}  \\
-e^{i\frac{2\pi}{3}}\sin\frac{\theta}{2}\cos\frac{\theta}{2} &
e^{i\frac{2\pi}{3}}\cos^{2}\frac{\theta}{2}
\end{array}\right), 
\label{T2_3term_p} \quad \quad
\end{eqnarray}
\end{subequations}
where $\tilde{c}=8kaqh_{1}/\tilde{y}$. In the above equations $\theta$, and 
$\tilde{c}$ are determined by the parameters $\left[ka,\omega/\Omega\right]$ 
calculated from the polarization condition given by Eq. (\ref{pol_cond}). Using 
Eqs. (\ref{T12_3term_p}), the polarized outputs $\left|\phi_{n}\right>$ can 
easily be determined as the eigenstates of the output density matrices 
$\varrho_{n}=\frac{1}{2}\tilde{T}_{n}(\tilde{T}_{n})^{\dagger}$ corresponding 
to the nonzero eigenvalues $\eta_{n}$.

\end{document}